\documentstyle[12pt]{article}

\def\be{\begin{equation}}
\def\ee{\end{equation}}
\def\ba{\begin{eqnarray}}
\def\ea{\end{eqnarray}}
\newcommand{\matel}[3]{\langle #1|#2|#3\rangle}
\newcommand{\ra}{\rightarrow}

\newlength{\dinwidth}
\newlength{\dinmargin}
\begin{document}
~~\\
\begin{flushright}
UND-HEP-BIG-99-05 \\ 
\end{flushright} 

\begin{center}
\begin{Large}
\begin{bf}
%
%
MEMO ON EXTRACTING $|V(cb)|$ 
AND $|V(ub)/V(cb)|$ FROM SEMILEPTONIC $B$ DECAYS
\end{bf}
\end{Large} \\ 
\vspace{5mm}
\begin{large}
%
%
I.I. Bigi \\
\end{large}
%
%
Physics Dept., University of Notre Dame du Lac\\
Notre Dame, IN 46556, U.S.A.\\
e-mail address: BIGI@UNDHEP.HEP.ND.EDU

\vspace*{2cm}
{\large \bf Abstract} 
\vspace*{.25cm} 
\end{center}
\noindent 
Heavy Quark Expansions for semileptonic decays of beauty hadrons 
are briefly reviewed. I analyze how $|V(cb)|$ can be extracted 
from the semileptonic width of $B$ mesons, the average 
semileptonic width of all weakly decaying beauty hadrons and from 
$B \ra l \nu D^*$ at zero recoil. Special attention is paid to 
present theoretical uncertainties (including correlations among 
them) and on how to reduce them in the future. Finally I will 
comment on theoretical uncertainties in $|V(ub)/V(cb)|$. 

\tableofcontents    

\section{Goal}
I will concentrate on semileptonic transitions since 
I do not see any realistic hope to extract $|V(cb)|$ from 
nonleptonic decays with competitive accuracy. I provide here an 
Executive 
Summary stating the relevant results, their theoretical 
uncertainties and how the latter can be reduced in the future. The 
derivations can be found in  
reviews like \cite{HQT,VARENNA} and in the original 
papers listed there. 

\section{Theoretical Technologies} 
\subsection{Fundamentals}

Heavy Quark Theory is implemented 
through the operator product expansion 
(OPE) in the form of a heavy quark expansion (HQE).  It allows to 
evaluate inclusive transition rates as an expansion in inverse 
powers of the heavy quark mass: 
\be 
\Gamma (B \ra l \nu X_q) = 
\frac{G_F^2 m_b^5}{192 \pi ^3} |V(qb)|^2 
\left[ c_3\langle B|O_3|B \rangle  + c_5
\frac{\langle B|O_5|B \rangle}{m_b^2} + 
c_6\frac{\langle B|O_6|B \rangle}{m_b^3} + 
{\cal O}(1/m_b^4) \right] 
\label{GENERAL} 
\ee 
where the $O_d$ denote local operators of 
(scale) dimension $d$. It is basically the same cast 
of operators -- albeit 
with different weights -- 
that appears in semileptonic, radiative and nonleptonic rates 
as well as distributions. 

The coefficients $c_d$ are 
calculated from short-distance physics; they contain the 
masses of the final state quarks from phase space etc. and powers 
of the strong coupling $\alpha _S$. The hadronic 
expectation values (HEV) of the operators $O_d$ encode the 
{\em nonperturbative} corrections. While we can identify these 
operators and their dimensions which then determine the power 
of $1/m_b$, in general we cannot (yet) 
compute their HEV's from first 
principles.  

For the HEV of the leading 
operator $O_3 = \bar bb$ one has 
\be 
\langle B|O_3|B \rangle /2M_B = 1 + {\cal O}(1/m_b^2) \; ; 
\ee 
it thus incorporates the parton model result which dominates 
asymptotically, i.e. for $m_b \ra \infty$. 

The most remarkable feature of Eq.(\ref{GENERAL}) is the 
{\em absence} of a contribution in order $1/m_b$. 
{\em Such a term is anathema in the OPE since there is no 
relevant operator of dimension four} 
\footnote{The operator $\bar Q \gamma \cdot \partial Q$ 
can be reduced to $m_Q \bar QQ$ by the equation of motion.}! 
This has an important consequence: 
\begin{itemize} 
\item 
With the leading nonperturbative corrections emerging in order 
$1/m_b^2$, they can amount to no more than typically 
several percent in beauty decays. More specifically they {\em 
reduce} the semileptonic $B$ width by close to 5 \%. That 
also means 
that evaluating the HEV's with no more than moderate accuracy 
-- say 20 \% -- already limits the overall uncertainties due to 
nonperturbative corrections to the 1\% level. 
\end{itemize} 
{\em Sum rules} \cite{OPTICAL} serve as 
an important theoretical tool: they yield well-defined 
relations between basic parameters of HQE and provide 
insights into how quark-hadron duality comes about.  

One warning should be stated explicitely: 
it would be {\em illegitimate} to merely replace 
{\em quark} masses in the OPE expressions by 
{\em hadron} masses. 
It can be shown \cite{FIVE} how sum rules lead to 
the emergence of quark phase space and the OPE 
nonperturbative corrections from the combination of 
hadronic phase space and bound state effects. 

\subsection{Determining the Size of Basic Parameters}

\subsubsection{Quark Masses}
Since the semileptonic width depends on the fifth power of the 
heavy quark mass, great care has to be applied in treating 
the latter. Masses like other parameters in a quantum field theory 
are not constants, but vary with the energy scale 
$\mu$ at which they are evaluated: $m_b(\mu )$. 
In the limit $\mu \ra 0$ the {\em pole} mass emerges;   
since -- due to a so-called renormalon singularity -- 
it suffers from an {\em irreducible} uncertainty 
$ \sim {\cal O}(1/m_Q)$ that is {\em larger} than the 
nonperturbative contributions one is evaluating, it is 
in principle inadequate for our purposes here.  
Such problems are avoided if one chooses a scale 
$\mu \geq 0.5$ GeV which shields $m_b(\mu )$ 
against renormalon 
singularities. It turns out that the natural choice for this 
scale is $\mu \sim m_b/5 \sim 1$ GeV. 
The $\overline{\rm MS}$ mass 
is ill-suited for describing {\em decay} 
(though not production) processes since it treats scales 
below its value in an unphysical way. Instead one 
adopts a more natural definition for the low 
scale running mass which leads to a {\em linear}  
dependance on $\mu$: 
\be 
\frac{dm_b(\mu )}{d\mu} = 
-c_m \frac{\alpha _S(\mu)}{\pi} + ... \; ; 
\ee 
the number $c_m$ specifies the concrete definition 
within this general class; reasonable choices are  
$c_m = 4/3$ or $16/9$. 

Since $\mu \sim 1$ GeV, $m_b(\mu )$ can conveniently 
be inferred from 
$e^+ e^- \ra b \bar b$ very close to threshold. That way one 
arrives at (for $c_m = 16/9$) 
\be
\label{22}
m_b(1 \; {\rm GeV}) = \left\{\begin{array}{ll}
4.56 \pm 0.06 \; {\rm GeV} & \cite{MELNIKOV1} \\ 
4.57 \pm 0.04 \; {\rm GeV} & \cite{HOANG}       \\ 
4.59 \pm 0.08 \; {\rm GeV} & \cite{BENEKE}                  
                         \end{array} \right.
\label{MB} 
\ee

The value of the mass difference $m_b - m_c$ 
can be inferred 
from the measured values of the spin averaged beauty 
and charm mesons: 
\ba 
m_b - m_c &=& \langle M_B\rangle - \langle M_D \rangle + 
\mu _{\pi}^2 \left( \frac{1}{2m_c} - \frac{1}{2m_b} \right) 
+ {\cal O}(1/m_{c,b}^2) 
\nonumber 
\\ 
&\simeq& 3.50 \; {\rm GeV}^2 + 40 \, {\rm MeV} 
\cdot \frac{\mu _{\pi}^2 - 0.5 \, {\rm GeV}^2}{0.1 \, {\rm GeV}^2} 
+ \Delta M_2 
\label{MBMCDIFF} 
\ea
where 
\be 
\langle M_{B[D]} \rangle = \frac{M_{B[D]} + 3 M_{B^*[D^*]}}{4}\; , 
\; \; \; |\Delta M_2| \leq 0.02 \, {\rm GeV}
\ee

\subsubsection{Hadronic Expectation Values}
There are two dimension-five operators, namely the Lorentz invariant 
chromomagnetic operator 
$\bar b \frac{i}{2} g_s\sigma \cdot G b$ 
and the kinetic energy operator $\bar b (i\vec D)^2b$ where 
$\vec D$ denotes the covariant derivative. The latter operator 
does not form a Lorentz scalar and therefore can enter only through 
the expansion of $\bar bb$. 
\begin{itemize} 
\item 
$\mu_G^2$: 

\noindent 
The size of the chromomagnetic HEV can be estimated 
through the hyperfine splitting 
\be 
\mu _G^2 \equiv \frac{1}{2M_B} 
\matel{B}{\bar b \frac{i}{2} g_s\sigma \cdot G b}{B} 
\simeq \frac{3}{4} \left( M_{B^*}^2 - M_B^2 \right) 
\pm 20 \% 
\simeq 0.36 \; {\rm GeV^2} \pm 20 \% \; , 
\ee 
yet it cannot automatically be equated with it. For the equality 
holds only asymptotically, namely in the limit $m_b \ra 
\infty$. At preasymptotic, i.e. finite values corrections of order 
$1/m_b$ (and higher) will 
arise. While it has been demonstrated in principle 
how those can be determined through the SV sum rules 
\cite{HQT}, this 
program has not been performed yet in a concrete way. 
This is a refinement that can be achieved in the future.

\item 
$\mu _{\pi}^2$:  

\noindent 
The quantity 
\be 
\mu _{\pi}^2 \equiv \frac{1}{2M_B} 
\matel{B}{\bar b (i\vec D)^2b}{B} 
\ee 
is related to the average kinetic energy of the $b$ quark 
moving inside the $B$ meson. 

While its precise value is not known yet, we can state 
the following: 
\begin{itemize}
\item 
An analysis based on QCD sum rules yields \cite{BALL}:
\be 
\mu _{\pi}^2 \simeq (0.5 \pm 0.15) \; {\rm (GeV)^2} \; . 
\ee 
\item 
In the literature $\mu _{\pi}^2$ is often equated with the 
HQET parameter $\lambda _1$ (or $-\lambda _1$). While 
the two quantities look superficially the same, they 
are {\em not} once quantum corrections are included. Those 
are essential for a self-consistent treatment. 
\item 
\be 
\mu _{\pi}^2 \geq \mu _G^2 
\ee 
holds as a field theoretical inequality 
\footnote{The dedicated reader should note that the exact 
definition of $\mu _{\pi}^2$ employed here is not identical 
to that of Ref.\cite{BALL}.}. 
\item 
Since $\mu _{\pi}^2 = \langle |\vec p_b|^2 \rangle$ one 
would be quite surprised if $\mu _{\pi}^2$ exceeded the 
bound $\mu _G^2 \simeq (0.6 \; {\rm GeV})^2$ considerably. 
\item 
Altogether a fairly conservative estimate is 
\be 
\mu _{\pi}^2 \simeq 0.5 \pm 0.2 \; {\rm GeV}^2 
\label{MUPI} 
\ee 

\end{itemize}
The remarks given above concerning preasymptotic 
corrections apply here as well. 
\end{itemize} 

\subsection{On Theoretical Uncertainties}
A few general comments on estimating uncertainties might be    
helpful: 
\begin{itemize}
\item 
The {\em conditio sine qua non} for the validity of the OPE result is quark-hadron duality. 
While it is expected to hold in the decays of 
beauty hadrons for $m_b \ra \infty$, there is a topical debate 
in the literature on its quantitative limitations for 
{\em finite} values 
of $m_b$ when applied to {\em nonleptonic} transitions. 
Yet here we are dealing with semileptonic decays which 
present a smoother dynamical stage. 

These issues can be analyzed in the 't Hooft model -- QCD in 1+1 dimensions with the 
number of 
colours $N_C \ra \infty$ -- which is an exactly soluble field theory; 
i.e., the spectrum of its boundstates and their wavefunctions are 
known. It automatically implements confinement. The validity 
of duality can be probed by comparing the OPE based result 
with the one obtained by summing over the hadronic 
resonances. Perfect matching has been found for 
semileptonic (as well as nonleptonic) widths. Furthermore 
duality violations can be modelled and were found to be 
suppressed by high powers of $1/m_Q$  
\cite{THOOFT,THOOFT2,THOOFT3}. 
\item 
In stating a theoretical error for a quantity I mean that the real 
value can lie almost anywhere in this range with basically 
equal probability rather than follow a Gaussian distribution. 
Furthermore my message is that I  
would be quite surprised if the real value would fall 
outside this range. Maybe one could call that a 
90 \% confidence level, but I do not see any way to be 
more quantitative. 
\item 
One of the most intriguing features of HQE 
is that they allow us to express a host of observables -- 
rates of nonleptonic, semileptonic and radiative decays, 
distributions etc. -- in terms of a handful of basic quantities, 
namely quark masses and expectation values of the 
leading operators as sketched above. 
This means, though, that {\em correlations} 
exist between the predictions for different observables 
which have to be accounted for. However, this is not always done 
in the literature. 
\item 
The just mentioned feature of  HQE can be exploited for 
important self-consistency checks by 
determining a basic quantity 
in two {\em systematically distinct} ways. If such a `redundant' extraction 
provides a self-consistent value, control has been established 
over the theoretical uncertainty; otherwise the need for 
further analysis has been revealed. 
\item 
For the HQE to make numerical sense, one needs 
\be 
m_Q \gg {\rm "typical" \; \; hadronic \; \; scales} 
\ee 
This requirement is certainly satisfied for {\em beauty} 
quarks for any reasonable definition of 
"typical" hadronic scales. Yet for {\em charm} quarks it 
makes a difference whether those are, say, 0.7 or 1 or 
1.2 GeV. At present we cannot decide this isssue with 
certainty. I will return to this point later on and 
illustrate it through an example.

\end{itemize}

\section{$\Gamma (B \ra l \nu X_c)$}

The basic expression is given by 
\be 
\Gamma _{SL}(B) = \frac{G_F^2m_b^5|V(cb)|^2}{192\pi ^3} 
\left[ z_0\left( 1 - \frac{\mu _{\pi}^2 - \mu _G^2}{2m_b^2}\right) 
- 2 \left( 1 - \frac{m_c^2}{m_b^2}\right) ^4 
\frac{\mu _G^2}{m_b^2} - 
\frac{2\alpha _S}{3\pi }z_0^{(1)} + ... \right] 
\label{MASTER} 
\ee 
where the omitted terms are higher-order perturbative corrections 
and/or power corrections; $z_0$, $z_0^{(1)}$ are {\em known} 
phase space factors depending on $m_c^2/m_b^2$.  The leading 
power corrections are controlled by the HEV's 
$ \mu _{\pi}^2$ and $\mu _G^2$. While the order $\alpha _S^2$ 
contributions have not been stated explicitely in 
Eq.(\ref{MASTER}), they are relevant and have been included in 
the numerical analysis.  

On the {\em conceptual} level the following has to be 
kept in mind: 
\begin{itemize}
\item 
While the expansion in Eq.(\ref{MASTER}) is formulated in 
inverse powers of $m_b$, the true expansion parameter is 
provided by the energy release. This is easily expressed for 
two complementary cases: for light final state quarks -- 
$m_c \ra 0$ -- the mass 
dependance is given by $m_b^5$, whereas for heavy ones 
-- $m_c \ra m_b$ -- it is $(m_b - m_c)^5$ 
\footnote{HQE are expected to provide a reliable 
treatment for $b\ra u$ as well as 
$b\ra c$ since $m_b$ as well as $m_b - m_c$ exceed ordinary 
hadronic scales $\bar \Lambda \sim \frac{1}{2}$ GeV.}. Intermediate 
scenarios can effectively be described in terms of powers of the difference 
$m_b - m_c$ and of $m_b$: 
$$
\Gamma _{SL}(b) \propto G_F^2 (m_b - m_c)^{5-p} m_b^p 
$$  
with $p \sim 2$. 

It should be noted that the energy release is quite a bit larger 
for $b \ra u$ than $b \ra c$ transitions. 
\item 
Although I did not make it explicit in the expressions above, 
the matrix elements depend on a normalization scale 
$\mu $. It emerges in the OPE through Wilson's 
prescription for separating long distance dynamics, which are  
lumped into the matrix elements, and short distance dynamics 
included in the c number coefficient functions of the 
operators in the OPE. The observable $\Gamma _{SL}(B)$ 
{\em in principle}   
cannot depend on this auxiliary scale since the 
$\mu$ dependance of the matrix elements obviously 
cancels against that of the coefficient functions in a complete 
calculation. {\em In practice} though one has to 
choose it judiciously for computational purposes: it has to satisfy 
\be 
\Lambda _{QCD} \ll \mu  \ll m_b
\ee 
to enable us to evaluate both the perturbative corrections and the 
HEV's. In particular that means that the (running) quark masses 
that enter in the coefficient functions 
have to be evaluated at the same scale $\mu $. 

I use $\mu \sim 1$ GeV for the results stated below. 

\item 
The normalization scale $\mu $ should not be confused 
with the argument in the strong coupling $\alpha _S$ entering 
in the last term of Eq.(\ref{MASTER}), which is often denoted by 
$\mu$ as well; yet I will refer to it as 
$\tilde \mu$. While my  
$\mu$ reflects the proper normalization of the various 
operators including those generating the nonperturbative 
contributions, the $\alpha _S(\tilde \mu )$ corrections 
are generated by  
the dynamics of high momenta of order  
$ \tilde \mu \gg \mu$. 

It turns out that the main impact of the radiative corrections 
has been accounted for once the quark masses have been 
evaluated at $\mu$ as stated above. The dependance 
on $\tilde \mu$ then becomes a secondary one. 
\footnote{ 
Having settled the scale at which $\alpha _S$ should be evaluated one still has to 
determine its size at that scale. Extrapolating down from values for $\alpha _S$ 
obtained at high scales like the $Z^0$ 
mass might not be quite straightforward. It has been suggested -- 
although not established -- that the `running' of the strong coupling 
is slowed down at lower scales due to nonperturbative effects.}  
\end{itemize}
{\em Numerically} one can state: 
\begin{itemize} 
\item 
The direct nonperturbative corrections of order $1/m_b^2$ are 
rather small and under control. 
\item 
The main challenge then consists of evaluating the 
{\em perturbative} expansion  
$$  
\Gamma (b \ra l \nu c) = \frac{G_F^2m_b^5(\mu )}
{192 \pi ^3} |V(cb)|^2 \left[ a_0(m_c/m_b) + 
a_1(m_c/m_b) \frac{\alpha _S(\tilde \mu )}{\pi} + \right. 
$$ 
\be 
\left.  a_2(m_c/m_b) \left( 
\frac{\alpha _S(\tilde \mu )}{\pi}\right)^2 + 
{\cal O}(\alpha _S^3) \right]  
\ee 
where the phase space corrections for $m_c \neq 0$ enter 
through the coefficients $a_i$. 
\item 
Any change in the normalization of the mass 
\be 
m_b(\mu ) \ra m_b(\mu + \Delta \mu) = 
m_b(\mu ) + \left[ k_1 \frac{\alpha _S}{\pi} + 
k_2 \left( \frac{\alpha _S}{\pi} \right) ^2 + ... \right] 
\Delta \mu 
\ee 
greatly affects the perturbative coefficients $a_i$, in particular 
because of the high power with which $m_b$ enters. 

If a calculation yields a 
large value for the second order coefficient 
$a_2$, one certainly has to be concerned that 
$a_3$ etc. could be sizeable as well. While this could signal that 
the perturbative corrections introduce a considerable 
uncertainty, it could also mean that these large coefficients are 
an {\em artefact} of choosing an unnatural scale for evaluating 
$m_b$. It has been shown \cite{FIVE} that the latter is the case 
here. Adopting $\mu \sim m_b/5$ effectively sums a sequence 
of higher-order contributions; the remaining coefficients are 
of order unity making the uncertainty due to perturbative 
corrections small. 

With the size of the quark masses of obvious importance they 
have to be defined properly and treated carefully, as mentioned 
above. 
\end{itemize} 
Equating the expression in Eq.(\ref{MASTER}) with the measured 
width one arrives at  
\ba 
|V(cb)|_{\Gamma _{SL}(B)} &=& 
0.0411 \sqrt{\frac{BR(B \ra l \nu X)}{0.105}} 
\sqrt{\frac{1.55 \, {\rm ps}}{\tau (B)}} 
\nonumber \\
&\times& \left( 1 - 0.025 \cdot 
\frac{\mu _{\pi}^2 - 0.5 \, {\rm GeV}^2}{0.2 \, {\rm GeV}^2} \right)  
\nonumber \\ 
&\times& \left( 1 \pm 0.01|_{m_b} \pm 0.01|_{pert}   
\pm 0.015|_{1/m_Q^3} \right) 
\label{VCBIN}
\ea
where the second and third lines refer to the theoretical 
uncertainties. I have taken into account here the 
correlations alluded to in the introduction. 
For the coefficient in the bracket in the second 
line reflects the fact that the size of $\mu _{\pi}^2$ 
affects also the value of $m_b - m_c$, see 
Eq.(\ref{MBMCDIFF}). The {\em remaining} sensitivity to $m_b$ 
enters through $\delta m_b$, where 
$\delta m_b/m_b \simeq 0.01$ has been assumed. The last term 
finally represents an estimate of power corrections 
beyond order $1/m_b^2$, including possible deviations from 
quark-hadron duality.   

The central value in Eq.(\ref{VCBIN}) is lower by 2\% 
than the 
one given in Ref.\cite{HQT} basically due to two reasons: 
(i) A value 
of about 4.51 GeV had been used in Ref.\cite{HQT} for the low scale 
{\em kinetic} $b$ quark mass, i.e. a number lower by 
about 1\% than what now the best available 
determinations yield. (ii) The non-BLM part of the 
${\cal O}(\alpha _S^2)$ contributions slightly increase the width. 
The final digit in the central value has to be taken with a 
grain of salt for the moment while further checks are being 
performed.  

Adding the theoretical 
uncertainties {\em linearly} rather than quadratically one   
arrives at 
\be 
\delta _{theor} |V(cb)|_{\Gamma _{SL}(B)} 
\simeq 6\%   \; \; \; {\rm at} \; {\rm present} 
\ee
I view such an error estimate as prudent rather than 
conservative. For there is reason for concern that our 
determinations of $m_b - m_c$ and of $m_b$ suffer from 
systematic short-comings. 
\begin{itemize}
\item 
Using Eq.(\ref{MBMCDIFF}) relies on an expansion in powers 
of $1/m_c$ to be numerically reliable. Yet we cannot rest 
assured of that. For the charm quark is 
not truly much larger than typical hadronic 
scales. 
Eq.(\ref{MUPI}) implies those scales are around 
$\sqrt{\mu _{\pi}^2} \sim 0.7$ GeV making an expansion 
in powers of $1/m_c$ meaningful. 

However this is not an 
ironclad conclusion. For it is not ruled out that the 
relevant hadronic scales are higher; after all we have only a 
lower bound on $\mu _{\pi}^2$. 
It is 
conceivable that both the central value of and 
the uncertainty in $m_b - m_c$ might differ significantly 
from the values stated above. Combining  
Eqs.(\ref{MBMCDIFF},\ref{MB}) leads to an uncomfortably small 
value of the charm quark mass. The value of 
$m_b - m_c$ thus contains a central theoretical 
uncertainty in evaluating $\Gamma (B \ra l \nu X_c)$. 
\item 
The difference $m_b - m_c$ can be extracted from 
moments of semileptonic decay spectra through methods 
that do not suffer the same 
dependance on $1/m_c$.  
\item 
All existing studies determine $m_b$ in very much the 
same way from $\Upsilon$ spectroscopy. More confidence in 
its value would be gained through extracting it by a different 
method. 
\end{itemize} 

The error estimate of Eq.(\ref{MASTER}) had been criticised in the 
past as unrealistically 
optimistic by authors that used pole masses or running masses 
evaluated at the high scale $m_b$. It was argued that the 
$m_b^5$ term suffers from a $\sim 10$\% uncertainty as do 
the radiative corrections; this would 
suggest an irreducible $\sim 20$\% theoretical uncertainty 
in the expression for the width. The point missed in these 
claims was the fact that these two sources of 
uncertainties are highly correlated as explained above.

I think that reducing this error down to 
\be 
\delta _{theor} |V(cb)|_{\Gamma _{SL}(B)} 
\simeq 2\%    
\ee
is 
an attainable goal over the next several years through refining our 
theoretical tools and calibrating them through experimental cross 
checks. 

However, I see it as quite unlikely 
that the uncertainties in $m_b$ and $m_b - m_c$ 
can be reduced to the desired level by theoretical 
means alone: a {\em systematically} different way for determining 
$m_b$ and $m_b - m_c$ can be obtained through a careful 
analysis of {\em moments} of semileptonic decay {\em spectra}. 
The necessary theoretical tool do exist and require just some 
polishing; the data can be obtained. 

\subsection{Criticism in the Literature}

The OPE treatment of the semileptonic decays of beauty hadrons 
has been criticised by Jin in several papers 
summarized in \cite{JIN} where he 
claims that 
\begin{enumerate} 
\item 
his ansatz is no less based on QCD than the OPE description; 
\item 
the OPE treatment misses relevant kinematical features; 
\item 
his reduced semileptonic width -- i.e. where the KM parameter 
has been factored out -- is larger than the OPE reduced width by 
a significant amount: 
\be 
\frac{1}{|V(qb)|^2} \Gamma ^{\rm Ref.\cite{JIN}} (B \ra l \nu X_q) > 
\frac{1}{|V(qb)|^2} \Gamma ^{\rm OPE} (B \ra l \nu X_q) \; ; 
\ee 
this would yield a lower value of $|V(qb)|$ obtained from the 
same data. 
\end{enumerate} 
While the last claim is factually correct the first two are not, 
which makes the last one irrelevant. 

Observable spectra are described through folding 
quark spectra with a (ligh cone) quark distribution  function. 
The actual choice of such a wave function which 
corresponds to {\em leading-}twist dynamics affects 
{\em spectra}; 
it {\em cannot}, however, have an impact on 
{\em integrated} rates as we are 
discussing here. 

In addition -- and probably most to the point -- his description 
is not only different, but inconsistent as expanding his expression 
for the semileptonic width reveals: 
\ba 
\Gamma ^{\rm Ref.\cite{JIN}} (B \ra l \nu X) &=& 
\Gamma (b \ra l \nu q) \cdot 
\left[ 1+ \frac{35}{6} \frac{\mu _{\pi}^2}{m_b^2} + ... \right] \\ 
\Gamma ^{\rm OPE} (B \ra l \nu X) &=& \Gamma (b \ra l \nu q) \cdot 
\left[ 1 - \frac{1}{2} \frac{\mu _{\pi}^2}{m_b^2} + ... \right]
\ea 
Note the difference in sign and magnitude -- 
$\frac{35}{6}/\frac{1}{2} \simeq 12$  -- of the 
$\mu _{\pi}^2/m_b^2$ terms in the two expressions. Without any 
calculation one can immediately see that the OPE result is correct and the 
other wrong. Since $\mu _{\pi}^2 = \langle |\vec p_b|^2 \rangle 
\simeq \langle m_b^2 |\vec v_b|^2 \rangle$ one realizes that the 
expression 
\be 
1 - \frac{1}{2} \frac{\mu _{\pi}^2}{m_b^2} \simeq 
1- \frac{1}{2} |\vec v_b|^2 \simeq \sqrt{1 - |\vec v_b|^2} 
\ee 
is nothing but the Lorentz time dilation factor! This kinematical 
effect has to be present. The expression of 
Ref. \cite{JIN} fails 
completely to reproduce the limiting case of a freely 
moving $b$ quark. This makes no sense in particular for an 
ansatz based on the parton model. 

Lastly the 't Hooft model is QCD in 1+1 dimensions with $N_C$, the 
number of colours, going to infinity. It retains many dynamical 
features of real QCD like confinement etc., 
yet can be solved exactly. One finds that the semileptonic $b\ra c$ 
as well as $b \ra u$ widths indeed match the OPE results 
\cite{THOOFT}. 

The question of quark versus hadron kinematics in HQE has been 
carefully studied since 1993 in several dedicated papers 
\cite{PRL,MOTION,OPTICAL,FIVE} and others more. 
It has been demonstrated how the full kinematical range can be 
covered, how the quark phase space and the OPE nonperturbative 
corrections emerge from the combination of hadronic phase space 
and bound state effects.

\section{$B \ra l \nu D^*$ at Zero Recoil}
\subsection{General Idea}
Having measured the lepton spectrum in $B \ra l \nu D^*$ one can 
extrapolate to the point of zero recoil to extract 
$|V(cb)F_{D^*}(0)|$. This extrapolation will introduce a theoretical 
uncertainty about which I have nothing new to add. 

Heavy quark symmetry tells us that 
\be 
{\rm lim} \, F_{D^*}(0) = 1    \; \; \; {\rm as} \; \; \; m_b \ra \infty 
\ee
must hold; i.e., we have 
$$  
F_{D^*}(0) = 1 + {\cal O}(\alpha _S/\pi ) + \delta ^A_{1/m^2} 
+ \delta ^A_{1/m^3} + ... 
$$ 
\be 
\delta ^A_{1/m^2} = {\cal O}(1/m_c^2 ) + {\cal O}(1/m_cm_b ) 
+ {\cal O}(1/m_b^2 ) 
\label{FDZERO}
\ee 
The theoretical challenge consists of calculating the perturbative 
as well as nonperturbative corrections to this exclusive process. 

The absence of nonperturbative corrections of order $1/m_{b,c}$ 
was first noted by Shifman and Voloshin 
\cite{SV}. It was later analyzed in 
more detailed by Luke \cite{LUKE} and is often referred to as Luke's 
theorem. 

Calculating the perturbative corrections provides us with a 
technical challenge where the last word has not been spoken yet; 
however they do not pose a conceptual problem. The conceptual 
challenge has come from the power corrections. 

Till 1994 the 
canonical claim had been that HQET allows to determine these 
contributions reliably and that they are quite small 
\be 
\delta ^A_{1/m^2} = ( -2 \pm 1)\% 
\ee  
leading to $F_{D^*}(0) \simeq 0.97$. 

 In 1994 the OPE-based heavy quark expansion was 
applied to this problem
\cite{OPTICAL,FD0}. Employing a 
(zero velocity) sum rule the analysis 
yielded $F_{D^*}(0) \simeq 0.9$.  
This value has now been widely adopted as the central value. 

\subsection{Power Corrections}

Heavy quark expansions allow to calculate the inclusive width for all 
channels generated by the axial current: 
$B \stackrel{j_{\mu}^5}{\longrightarrow} X_A$. The hadronic 
final state $X_A$ has three components, namely 
\begin{itemize} 
\item 
$D^*$ as the lowest state; 
\item 
higher excitations like $D^{**}$, etc. and 
\item 
a quasi-continuum formed by $D+\pi 's$ configurations where 
the distinction between this and the preceding component is 
fluid. 

\end{itemize} 
Turning the argument around 
$B \ra D^*$ is given as the difference between the inclusive 
zero recoil width 
$B \ra X_A$ and the higher excitations including the 
continuum. The 
formfactor $F_{D^*}$ can thus be expressed through three types of 
contributions: 
\begin{enumerate}
\item 
those coming from local operators emerging in a 
$1/m_Q$ expansion, as characterized by $\mu _{\pi}^2$, 
$\mu _G^2$ and terms of order $1/m_Q^3$ (and higher). 
One should 
note here that there are three variants of $1/m^2$ 
terms, namely $1/m_c^2$, $1/(m_c m_b)$ and 
$1/m_b^2$. Obviously the first one will usually 
dominate. 
\item 
{\em nonlocal} contributions from higher excitations;  
\item 
perturbative ones. 
\end{enumerate}  
Combining them I obtain  
\ba 
F_{D^*} (0) &\simeq& 0.89 - 
0.026 \cdot \frac{\mu _{\pi}^2 - 0.5 \, {\rm GeV^2}}{0.2 \, 
{\rm GeV^2}} \pm 0.02|_{excit} \pm 0.01 |_{pert} 
\pm 0.025_{1/m^3} 
\nonumber 
\\ 
&\simeq& 0.89 \pm 0.08 |_{theor} 
\label{ERROREX} 
\ea 
The central value has been lowered since a more careful 
analysis of the perturbative corrections that partially 
accounts for $1/m_Q^3$ terms suggests a smaller short 
distance renormalization factor \cite{KOLYAPRIVATE}. 

A few comments will help to gain a proper perspective 
of these findings: 
\begin{itemize}
\item 
One should note that the impact of $\mu _{\pi}^2$ 
on $F_{D^*}(0)$ and on $\Gamma _{SL}(B)$ is 
{\em anticorrelated}: while a larger value of $\mu _{\pi}^2$ 
{\em reduces} $F_{D^*}(0)$ and thus leads to a 
{\em larger} value 
of $|V(cb)|$ to be obtained from the observable 
$|F_{D^*}(0) V(cb)|$, it {\em enhances} $\Gamma _{SL}(B)$ 
through a larger value for $m_b - m_c$ and thus 
{\em reduces} $|V(cb)|_{\Gamma _{SL}(B)}$, 
see Eqs.(\ref{VCBIN},\ref{ERROREX}). 
\item 
The value of 
$\mu _{\pi}^2$ represents a major common source of the 
uncertainty in both cases; however  
the origin is very different: 
\begin{itemize}
\item 
In $\Gamma _{SL}(B)$ it mainly reflects how we infer the value 
of $m_b - m_c$. Once the latter has been determined differently, 
namely 
from spectra in semileptonic decays, the 
impact of $\mu _{\pi}^2$ will largely fade as far as 
$\Gamma _{SL}(B)$ is concerned. 
\item 
For $F_{D^*}(0)$ on the other hand $\mu _{\pi}^2$ forms an 
essential   
part of our evaluation. 
\end{itemize} 
\item 
The leading nonperturbative corrections 
to $F_{D^*} (0)$ are controlled by a 
$1/m_c^2$ term. A deviation of $F_{D^*} (0)$ from unity 
by about 10\% is then much more reasonable then the 
originally 
claimed 2\%. Yet this raises also the legitimate concern 
whether such an expansion is numerically reliable because 
of the charm quark mass being only moderately large. While 
I have expressed this worry also in the discussion of the 
total semileptonic width, it is much more serious here: 
\begin{itemize}
\item 
The $1/m_c$ expansion enters the treatment of 
$\Gamma _{SL}(B)$ only in a secondary way, namely as a tool 
to determine $m_b - m_c$. As stated before, that mass 
difference can be extracted from spectra in semileptonic 
$B$ decays in ways that do {\em not} depend on a 
$1/m_c$ expansion. 
\item 
In $F_{D^*}(0)$ on the other hand the $1/m_c$ expansion forms 
an integral part of the analysis and no good way has been found 
to determine these power corrections in a more reliable 
way. 
\end{itemize} 
\item 
Estimates of the uncertainties in the $1/m^3$ corrections 
and of the contributions from the higher excitations do not have 
a firm foundation yet. 
\item 
At present I see little basis for adding the 
theoretical uncertainties in quadrature. 
\end{itemize} 
Some moderate reduction in some of the uncertainties should be 
achievable, in particular concerning $\mu _{\pi}^2$; 
modelling the excitations might also provide some 
useful handles. Yet in contrast to the situation with the total 
semileptonic $B$ width I do not see how the theoretical 
uncertainty in $F_{D^*}(0)$ can be reduced significantly, i.e. be cut in 
half. For that to come about  
a genuine breakthrough had to happen.

\section{Using $ \tau (b)$ \& BR$(b \ra l \nu X)$}
Rather than determining the lifetimes and semileptonic 
branching ratios of {\em specific} beauty hadrons one can 
measure these observables 
{\em averaged} over the beauty hadrons 
present in the sample. This pedestrian implementation 
of quark-hadron duality will certainly minimize the 
statistical errors. The question is how great a price one pays 
in systematic uncertainties. 

The average semileptonic branching ratio of 
beauty hadrons in a given sample is expressed by 
\be 
\langle BR_{SL}(b) \rangle = 
w_{B_d} \frac{\Gamma _{SL}(B_d)}{\Gamma (B_d)} + 
w_{B^+} \frac{\Gamma _{SL}(B^+)}{\Gamma (B^+)} + 
w_{B_s} \frac{\Gamma _{SL}(B_s)}{\Gamma (B_s)} + 
w_{\Lambda _b} \frac{\Gamma _{SL}(\Lambda _b)}
{\Gamma (\Lambda _b)} 
\label{AVERAGEBR} 
\ee
with the $w_{H_b}$ denoting the relative weights of the various 
beauty hadrons in the data sample under study; $\Lambda _b$ is 
used in a generic way to 
include other beauty baryons $\Xi _b^{0,-}$. 

One predicts that the semileptonic widths of the 
beauty {\em mesons} differ by less than a percent 
\be 
\Gamma _{SL}(B_d) \simeq \Gamma _{SL}(B^+) \simeq 
\Gamma _{SL}(B_s) \equiv \Gamma _{SL}(B)   
\ee 
Using the constraint 
\be 
w_{B_d} + w_{B^+} + w_{B_s} + w_{\Lambda _b}  = 1 
\ee 
one arrives at 
\be 
\langle BR_{SL}(b) \rangle \simeq \Gamma _{SL}(B) 
\langle \tau (b) \rangle 
\left[ 1 + w_{\Lambda _b} \cdot 
\frac{\tau (\Lambda _b)}{\langle \tau (b) \rangle } 
\left( \frac{\Gamma _{SL}(\Lambda _b)}{\Gamma _{SL}(B)} -1 
\right) \right] 
\ee 
where $\langle \tau (b) \rangle $ denotes the average lifetime of 
beauty hadrons in the sample. It has been measured with 
considerable accuracy at LEP where the `beauty cocktail' 
is described by 
\be 
w_{B^+} = w_{B_d} = 0.395 ^{+0.013}_{-0.014} \, , \; 
w_{B_s} = 0.108 \pm 0.014 \, , \;  
w_{\Lambda _b} = 0.102 ^{+0.023}_{-0.021}
\ee 
I do not claim a quantitative understanding of beauty baryon 
lifetimes -- nor do I need to. For the second factor in the 
square bracket will not amount to more than a one percent 
effect or so, since $w_{\Lambda _b} \simeq 0.1 \pm 0.02$ 
and 
$|\Gamma _{SL}(\Lambda _b)/\Gamma _{SL}(B) - 1| \leq 
0.2$ represents a generous bound for {\em semileptonic} 
widths. 

The measurement of the average semileptonic branching ratio 
and beauty lifetime thus amounts to a measurement of 
$\Gamma _{SL}(B)$ with {\em no} additonal 
uncertainties of numerical significance. The same is likewise 
true for $\langle BR(b \ra l \nu X_u)\rangle$ and 
$\Gamma (B \ra l \nu X_u)$.   

\section{Theoretical Uncertainty in $|V(ub)/V(cb)|$}

\subsection{$|V(ub)|$}

$|V(ub)|$ can be determined from the measured value for 
$\Gamma (B \ra l \nu X_u)$ again using HQE technology 
(see \cite{URIVUB} for a summary): 
\ba 
|V(ub)|_{B \ra l \nu X_u} = 0.00442 &\cdot & 
\left(  \frac{{\rm BR}(B \ra l \nu X_u)}{0.002}\right) ^{1/2} 
\left( \frac{1.55 \; {\rm ps}}{\tau _B}\right) ^{1/2} 
\nonumber \\ 
&\cdot& 
\left( 1 \pm 0.01|_{pert} \pm 0.018|_{1/m_b^3} \pm 0.035|_{m_b} \right) 
\label{VUB} 
\ea 

As far as the angles of the KM unitarity triangle are concerned 
which control CP asymmetries in $B$ decays, one is interested 
more in $|V(ub)/V(cb)|$ than $|V(ub)|$ itself. It is then quite relevant to analyze whether some theoretical uncertainties drop out from this ratio. 

\subsection{$|V(ub)/V(cb)|$}

Most of the theoretical uncertainties in 
$|V(cb)|_{\Gamma _{SL}(B)}$ and 
$|V(ub)|_{B \ra l \nu X_u}$, Eqs.(\ref{VCBIN},
\ref{VUB}), are {\em independant} or only mildly correlated: 
\begin{itemize}
\item 
Whereas the value of $\mu _{\pi}^2$ has great impact on  
$|V(cb)|_{\Gamma _{SL}(B)}$ as 
yardstick for $m_b - m_c$, it affects 
$|V(ub)|_{B \ra l \nu X_u}$ very 
little. 
\item 
The probably leading effect in ${\cal O}(1/m_b^3)$, namely 
the weak annihilation contribution to semileptonic 
$B^+$ decays, does not affect $|V(cb)|_{\Gamma _{SL}(B)}$. 
\item 
While a priori the perturbative uncertainties could be highly 
correlated for $b\ra c$ and $b \ra u$, it turns out to be otherwise. 
\end{itemize} 
I will add the $1/m_b^3$ uncertainty in 
$|V(ub)|_{B \ra l \nu X_u}$ and 
$|V(cb)|_{\Gamma _{SL}(B)}$ in 
quadrature; likewise for the perturbative one. 

One uncertainty is significantly correlated: 
\begin{itemize}
\item 
The value of $m_b(1\; {\rm GeV})$ is obviously an 
important input in both cases, although less so in 
$b \ra c$ where $\Gamma (B \ra l \nu X_c)  
\propto (m_b - m_c)^{5-p} m_b^p$ 
with $p \sim 2$ than in 
$b\ra u$ with 
$\Gamma (B \ra l \nu X_u)
\propto  m_b^5$. 
\end{itemize}
Hence I infer the following theoretical uncertainties for 
$|V(ub)|_{B \ra l \nu X_u}/|V(cb)|_{\Gamma _{SL}(B)}$: 
\be 
\delta _{theor} (|V(ub)|_{B \ra l \nu X_u}/
|V(cb)|_{\Gamma _{SL}(B)}) 
= \pm 0.02 |_{m_b} \pm 0.025|_{\mu _{\pi}^2} 
\pm 0.018|_{pert} \pm 0.03|_{1/m_b^3} 
\label{VUBVCBIN} 
\ee 
Adding these errors linearly might be overly conservative; 
adding them {\em in quadrature} one 
arrives at an overall uncertainty of 4 - 5\%. 

The situation with respect to 
$|V(ub)|_{B \ra l \nu X_u}/|V(cb)|_{B\ra D^*}$ 
is simpler in the sense that there are hardly any correlations 
in the theoretical uncertainties. Adopting the same strategy 
as just sketched, I estimate: 
\be 
\delta _{theor} (|V(ub)|_{B \ra l \nu X_u}/|V(cb)|_{B\ra D^*}) 
= 0.025|_{\mu _{\pi}^2} \pm 0.02|_{excit} 
\pm 0.035 |_{m_b}  \pm 0.04|_{1/m_b^3}  
\pm 0.015|_{pert} 
\ee 
representing an overall uncertainty of 6 - 7 \% 
when added in 
quadrature (adding them linearly might be even less 
reasonable than for Eq.(\ref{VUBVCBIN})).

\section{Summary and Conclusions}

The significant conclusions from this discussion 
can be summarized as follows: 
\begin{itemize}
\item 
The KM parameter $|V(cb)|$ can at present 
be extracted  
\begin{itemize} 
\item 
from the total 
semileptonic width, 
\item 
from $B \ra l \nu D^*$ at zero recoil and 
\item 
from the `average' beauty semileptonic branching 
ratio and lifetime 
\end{itemize} 
with about a 6\% {\em theoretical} accuracy. 
\item 
$|V(ub)|$ can be obtained from $\Gamma (B \ra l \nu X_u)$ 
or $\langle {\rm BR}(b \ra l \nu u)\rangle$ in conjunction with 
$\langle \tau (b) \rangle$ with about a 7-8 \% theoretical 
accuracy. 
\item 
The ratio $|V(ub)/V(cb)|$ can the be determined with an 
uncertainty of better than 5\%. 
\item 
The very attractive fundamental feature 
of Heavy Quark Theory that many different obserables 
are related to each other -- sometimes in subtle ways -- 
means that there are numerous correlations and that any 
change in one of the basic quantities implies changes in several 
observables. This has to be taken into account. 
\item 
Presently it is prudent to combine theoretical errors 
{\em linearly} as it was mostly done above: 
theoretical errors in general do not follow 
a Gaussian distribution; correlations are not always 
manifest; expansions in $1/m_c$ might not be numerically 
reliable; the value of the crucial quantity $m_b$ has so far 
been inferred from a single reaction only, etc. 
\item 
This problem can be overcome, though. Again using the very fact 
of few quantities controlling several a priori independant 
observables one will be able -- in due time -- to extract 
the values of these quantities in different ways. If these 
redundant extractions indeed agree one has established 
a new level of theoretical control and can add also theoretical 
uncertainties in quadrature. 
\item 
We can map out the strategy for reducing the theoretical error 
in $\Gamma _{SL}(B)$ to 2\%. It is hard to see how the same 
improvement can be achieved for the description of 
$B \ra l \nu D^*$. 

\end{itemize}

\vspace*{.2cm} 

{\bf Acknowledgements:} 

I am grateful to my collaborators M. Shifman and N. Uraltsev 
for making valuable suggestions concerning this text. 
I truly enjoyed the informal workshop organized by 
M. Battaglia, P. Henrard and other members of the 
LEP Heavy Flavour Steering Group and have benefitted 
greatly from it. 
This 
work has been supported by the National Science 
Foundation under grant number PHY96-05080.

\end{document}